\documentclass[a4paper,aps,onecolumn,nofootinbib]{revtex4}
\RequirePackage[colorlinks,hyperindex]{hyperref}
\RequirePackage[english]{babel}
\RequirePackage[latin1]{inputenc}
\RequirePackage[T1]{fontenc}
\RequirePackage{mathrsfs}
\RequirePackage{amsmath}
\RequirePackage{amssymb}
\RequirePackage{amsbsy}
\RequirePackage{color}
\RequirePackage{bm}
\hypersetup{colorlinks=true,breaklinks=true,urlcolor=blue,linkcolor=red}
\begin{document}
\title{Analytic exact solutions to the nonlinear Dirac equation}
\author{Luca Fabbri$^{c}$\!\!\! $^{\hbar}$\!\!\! $^{G}$\footnote{luca.fabbri@unige.it},
Roberto Cianci$^{c}$\!\!\! $^{\hbar}$\!\!\! $^{G}$\footnote{roberto.cianci@unige.it}}
\affiliation{$^{c}$DIME, Universit\`{a} di Genova, Via all'Opera Pia 15, 16145 Genova, ITALY\\
$^{\hbar}$INFN, Sezione di Genova, Via Dodecaneso 33, 16146 Genova, ITALY\\
$^{G}$GNFM, Istituto Nazionale di Alta Matematica, P.le Aldo Moro 5, 00185 Roma, ITALY}
\date{\today}
\begin{abstract}
We present analytic exact solutions to the nonlinear Dirac equation: they display a ring singularity for the Nambu--Jona-Lasinio nonlinearity and a shell singularity for the Soler nonlinearity. For both cases, the size of the singular region is of the order of the Compton length. General comments about square-integrability will be addressed in terms of a solution to the linearized system.
\end{abstract}
\maketitle
\section{Introduction}
The 1950s is the decade when physicists started to pay attention to nonlinear dynamics in quantum mechanics \cite{RevModPhys.29.269}.

Interesting applications were found about condensates coming straight from the nonrelativistic theory \cite{Gr, Pit}. Others were works involving general theoretical considerations on relativistic field theories in $(1\!+\!1)$-dimensional cases \cite{Th, Gr-Nev}.

In both situations, the nature of the nonlinearity is always that of a quartic potential, uniquely determined up to some scaling factor. In the physical $(1\!+\!3)$-dimensional spacetime, instead, the nonlinear terms in the Lagrangian are of two types, one given by the squared of the bilinear scalar and one given by the squared of the bilinear pseudo-scalar.\footnote{It is in fact possible to demonstrate, by employing Fierz re-arrangements, that for a Dirac spinor $\psi$ and its adjoint $\overline{\psi}$, all terms formed by four fields can always be reduced to either $(\overline{\psi}\psi)(\overline{\psi}\psi)$ or $(i\overline{\psi}\boldsymbol{\pi}\psi)(i\overline{\psi}\boldsymbol{\pi}\psi)$ alone. Here the matrix $\boldsymbol{\pi}$ is our notation for the pseudo-matrix.}

Consequently, there are two types of models: the first, considering purely scalar (spinless) self-interactions, is called \emph{Soler model} \cite{PhysRevD.1.2766}, \cite{Cazenave1,Cazenave2,Esteban}; the second, considering equally the scalar and the pseudo-scalar (chiral) self-interactions, is the \emph{Nambu--Jona-Lasinio model} \cite{N-JL}. There are also intermediate models, in which both nonlinearities are accounted for, but not with the same strength \cite{PhysRev.83.326,PhysRev.103.1571}: these models are capable of interpolating both Soler and Nambu--Jona-Lasinio models, with the Soler and the Nambu--Jona-Lasinio models being recovered as two specific limiting situations.

From a physical point of view, the N-JL model is interesting because it can be justified as the theory that we obtain when the Dirac equation in interaction with a propagating torsion is taken in the approximation in which torsion is considered to have a large mass \cite{Fabbri:2017rjf}. On the other hand, the S model might be justified as the theory that we obtain when the Dirac equation in interaction with the Higgs field is taken in the approximation in which the Higgs boson is considered to have a large mass \cite{Fabbri:2018css}. Therefore, the N-JL and S models are those with physical meaning.

However, for either model, no analytic exact solution is known. For the S model, Soler himself provided solutions, though only numerical ones \cite{PhysRevD.1.2766}. Others maintained the treatment mathematical discussing the existence of solutions, but they never actually showed one explicitly \cite{Cazenave1,Cazenave2,Esteban}. In \cite{PhysRev.83.326} and \cite{PhysRev.103.1571} Finkelstein and co-workers studied a spectrum of nonlinearities, but explicit results were given only for the S model. In this paper we show a method with which exact analytic solutions can be found for both models, explicitly providing a solution for each.
\section{Geometry of spinors}
In the following, the Clifford matrices $\boldsymbol{\gamma}_{a}$ verify $\left\{\boldsymbol{\gamma}_{a},\boldsymbol{\gamma}_{b}\right\}\!=\!2\eta_{ab}\mathbb{I}$ where $\eta_{ab}$ is the Minkowskian metric. Hence, sigma matrices $\boldsymbol{\sigma}_{ab}$ defined as $\boldsymbol{\sigma}_{ab}\!=\!\left[\boldsymbol{\gamma}_{a},\boldsymbol{\gamma}_{b}\right]/4$ are generators of spinor transformations. And finally, the parity-odd matrix $\boldsymbol{\pi}$ is implicitly defined by $2i\boldsymbol{\sigma}_{ab}\!=\!\varepsilon_{abcd}\boldsymbol{\pi}\boldsymbol{\sigma}^{cd}$ where $\varepsilon_{abcd}$ is the completely antisymmetric pseudo-tensor of Levi-Civita.

With a spinor and its adjoint $\overline{\psi}\!=\!\psi^{\dagger}\boldsymbol{\gamma}^{0}$ we can define the bilinear quantities
\begin{gather}
i\overline{\psi}\boldsymbol{\pi}\psi\!=\!\Theta\ \ \ \ \ \ \ \ \ \ \ \
\overline{\psi}\psi\!=\!\Phi\\
\overline{\psi}\boldsymbol{\gamma}^{a}\boldsymbol{\pi}\psi\!=\!S^{a}\ \ \ \ \ \ \ \ \ \ \ \
\overline{\psi}\boldsymbol{\gamma}^{a}\psi\!=\!U^{a}
\end{gather}
are all real tensors, verifying
\begin{gather}
U_{a}U^{a}\!=\!-S_{a}S^{a}\!=\!\Theta^{2}\!+\!\Phi^{2}\label{norm2}\\
U_{a}S^{a}\!=\!0\label{orthogonal2}
\end{gather}
called Fierz re-arrangements, all of which being geometric identities.\footnote{These are only a few of the Fierz re-arrangements: for a complete list and some of their demonstrations see \cite{Cavalcanti:2020obq, Ablamowicz:2014rpa}.}

For the soldering between the Lorentz structure with the manifold we use $\xi^{a}_{\nu}$ and $\xi_{k}^{\sigma}$ known as tetrad and co-tetrad fields. From them, the spin connection $C_{ab\mu}$ can be defined, so that one can write the covariant derivative of spinorial fields in general backgrounds. These concepts are part of basic courses of differential geometry.

For the dynamics, we consider the Dirac equation in the two cases we have mentioned before, in the introduction:
\begin{enumerate}
\item the torsion-spinor interaction -- in this case, the spinor field is taken to interact with the torsion of the spacetime according to the Lagrangian given by
\begin{eqnarray}
\mathscr{L}\!=\!\frac{1}{4}(\partial W)^{2}\!-\!\frac{1}{2}M^{2}W^{2}
\!-\!\frac{i}{2}(\overline{\psi}\boldsymbol{\gamma}^{\mu}\boldsymbol{\nabla}_{\mu}\psi
\!-\!\boldsymbol{\nabla}_{\mu}\overline{\psi}\boldsymbol{\gamma}^{\mu}\psi)
\!+\!XW_{\mu}\overline{\psi}\boldsymbol{\gamma}^{\mu}\boldsymbol{\pi}\psi\!+\!m\overline{\psi}\psi
\end{eqnarray}
where $(\partial W)_{\alpha\nu}\!=\!\nabla_{[\alpha}W_{\nu]}$ is the curl of the torsion $W_{\nu}$, $M$ is the torsion mass, and $X$ the torsion-spinor coupling constant. Interested readers can see the details in \cite{Fabbri:2024avj}. Variation furnishes the field equations
\begin{eqnarray}
\nabla_{\rho}(\partial W)^{\rho\mu}\!+\!M^{2}W^{\mu}
\!=\!X\overline{\psi}\boldsymbol{\gamma}^{\mu}\boldsymbol{\pi}\psi\label{torsionfe}
\end{eqnarray}
for torsion and
\begin{eqnarray}
i\boldsymbol{\gamma}^{\mu}\boldsymbol{\nabla}_{\mu}\psi
\!-\!XW_{\mu}\boldsymbol{\gamma}^{\mu}\boldsymbol{\pi}\psi\!-\!m\psi\!=\!0\label{spinorfe}
\end{eqnarray}
for the spinorial field. In effective approximation of torsion \cite{Fabbri:2017rjf}, terms in $(\partial W)_{\alpha\nu}$ are negligible if compared to those in $W_{\nu}$, making (\ref{torsionfe}) reduce to
\begin{eqnarray}
M^{2}W^{\mu}\!=\!X\overline{\psi}\boldsymbol{\gamma}^{\mu}\boldsymbol{\pi}\psi
\end{eqnarray}
where now torsion can be substituted in (\ref{spinorfe}) yielding
\begin{eqnarray}
i\boldsymbol{\gamma}^{\mu}\boldsymbol{\nabla}_{\mu}\psi
\!+\!\frac{X^{2}}{M^{2}}(\Phi\mathbb{I}\!+\!i\Theta\boldsymbol{\pi})\psi\!-\!m\psi\!=\!0
\end{eqnarray}
in which Fierz identities were used. With the normalization for which $X^{2}/M^{2}\!=\!1/4$ we get
\begin{eqnarray}
i\boldsymbol{\gamma}^{\mu}\boldsymbol{\nabla}_{\mu}\psi
\!+\!\frac{1}{4}(\Phi\mathbb{I}\!+\!i\Theta\boldsymbol{\pi})\psi\!-\!m\psi\!=\!0
\label{D:N-JL}
\end{eqnarray}
called Nambu--Jona-Lasinio equation \cite{N-JL}. The positive constant of the nonlinearity ensures its attractiveness;

\item the Higgs-spinor interaction -- this case is the scalar-fermion sector of the Standard Model with Lagrangian
\begin{eqnarray}
\mathscr{L}\!=\!-\frac{1}{2}(\nabla H)^{2}\!+\!\frac{1}{2}M^{2}H^{2}
\!-\!\frac{i}{2}(\overline{\psi}\boldsymbol{\gamma}^{\mu}\boldsymbol{\nabla}_{\mu}\psi
\!-\!\boldsymbol{\nabla}_{\mu}\overline{\psi}\boldsymbol{\gamma}^{\mu}\psi)
\!+\!YH\overline{\psi}\psi\!+\!m\overline{\psi}\psi
\end{eqnarray}
where now $M$ is the Higgs mass, and $Y$ the scalar-fermion coupling constant. This Lagrangian is the one of the Standard Model after symmetry breaking and in which all higher-order self-interactions of the Higgs have been neglected. Variation furnishes the field equations
\begin{eqnarray}
\nabla^{2}H\!+\!M^{2}H\!=\!-Y\overline{\psi}\psi\label{Higgsfe}
\end{eqnarray}
for the scalar and
\begin{eqnarray}
i\boldsymbol{\gamma}^{\mu}\boldsymbol{\nabla}_{\mu}\psi
\!-\!YH\psi\!-\!m\psi\!=\!0\label{spinorfe'}
\end{eqnarray}
for the fermion. In effective approximation of the Higgs field \cite{Fabbri:2018css}, (\ref{Higgsfe}) reduces to
\begin{eqnarray}
M^{2}H\!=\!-Y\overline{\psi}\psi
\end{eqnarray}
and having the Higgs field substituted in (\ref{spinorfe'}) produces
\begin{eqnarray}
i\boldsymbol{\gamma}^{\mu}\boldsymbol{\nabla}_{\mu}\psi
\!+\!\frac{Y^{2}}{M^{2}}\Phi\psi\!-\!m\psi\!=\!0
\end{eqnarray}
similarly to what we have seen for torsion. With the normalization $Y^{2}/M^{2}\!=\!1/4$ we get
\begin{eqnarray}
i\boldsymbol{\gamma}^{\mu}\boldsymbol{\nabla}_{\mu}\psi
\!+\!\frac{1}{4}\Phi\psi\!-\!m\psi\!=\!0
\label{D:S}
\end{eqnarray}
called Soler equation \cite{PhysRevD.1.2766}. As for the Nambu--Jona-Lasinio equation, positive nonlinearity indicates attractiveness.
\end{enumerate}

The Nambu--Jona-Lasinio and Soler nonlinear equations can be merged together into the single
\begin{eqnarray}
i\boldsymbol{\gamma}^{\mu}\boldsymbol{\nabla}_{\mu}\psi
\!+\!\frac{1}{4}(\Phi\mathbb{I}\!+\!ip\Theta\boldsymbol{\pi})\psi\!-\!m\psi\!=\!0
\label{D:merge}
\end{eqnarray}
reducing for $p\!=\!1$ to (\ref{D:N-JL}) and for $p\!=\!0$ to (\ref{D:S}). This form can account for those intermediate situations in which both the axial-vector and the scalar nonlinearities are present \cite{PhysRev.83.326,PhysRev.103.1571}. And when looking for solutions, equation (\ref{D:merge}) has the advantage of allowing us to find explicit solutions for both cases in a single form.

To search for analytic exact solutions, we first need to introduce the concept of polar decomposition.
\section{The polar form}
It is possible to demonstrate \cite{jl} that, if not both $\Theta$ and $\Phi$ are identically zero, then spinors can always be written in the so-called polar form, given (when in chiral representation) according to
\begin{gather}
\psi\!=\!\phi e^{-\frac{i}{2}\beta\boldsymbol{\pi}}
\boldsymbol{L}^{-1}\left(\!\begin{tabular}{c}
$1$\\
$0$\\
$1$\\
$0$
\end{tabular}\!\right)
\label{spinor}
\end{gather}
where the module $\phi$ and the chiral angle $\beta$ are given by
\begin{gather}
\Theta\!=\!2\phi^{2}\sin{\beta}\ \ \ \ \ \ \ \ \ \ \ \ \ \ \ \  \Phi\!=\!2\phi^{2}\cos{\beta}
\end{gather}
and where $\boldsymbol{L}$ is the spinor transformation needed to take the spinor into its rest-frame and spin-eigenstate. Also
\begin{gather}
S^{a}\!=\!2\phi^{2}s^{a}\ \ \ \ \ \ \ \ \ \ \ \ \ \ \ \  U^{a}\!=\!2\phi^{2}u^{a}
\label{SU}
\end{gather}
in terms of the spin $s^{a}$ and the velocity $u^{a}$ verifying $u_{a}u^{a}\!=\!-s_{a}s^{a}\!=\!1$ and $u_{a}s^{a}\!=\!0$ as reduced Fierz re-arrangements.

At differential level, it is possible to prove \cite{Fabbri:2024avj} that, in the most general case, the covariant derivative of the spinor field can be written in polar form as
\begin{eqnarray}
\boldsymbol{\nabla}_{\mu}\psi\!=\!\left(\nabla_{\mu}\ln{\phi}\mathbb{I}
\!-\!\frac{i}{2}\nabla_{\mu}\beta\boldsymbol{\pi}
\!-\!iP_{\mu}\mathbb{I}\!-\!\frac{1}{2}R_{ij\mu}\boldsymbol{\sigma}^{ij}\right)\psi
\label{decspinder}
\end{eqnarray}
in terms of the gauge-invariant vector $P_{\mu}$ called momentum and of the real antisymmetric tensor $R_{ab\mu}\!=\!-R_{ba\mu}$ called tensorial connection. We have
\begin{gather}
\nabla_{\mu}s_{i}\!=\!s^{j}R_{ji\mu}\ \ \ \ \ \ \ \
\ \ \ \ \ \ \ \ \nabla_{\mu}u_{i}\!=\!u^{j}R_{ji\mu}\label{ds-du}
\end{gather}
as geometric identities. We also have that
\begin{gather}
R^{i}_{\phantom{i}j\mu\nu}\!=\!-(\nabla_{\mu}R^{i}_{\phantom{i}j\nu}
\!-\!\nabla_{\nu}R^{i}_{\phantom{i}j\mu}
\!+\!R^{i}_{\phantom{i}k\mu}R^{k}_{\phantom{k}j\nu}
\!-\!R^{i}_{\phantom{i}k\nu}R^{k}_{\phantom{k}j\mu})\label{Rie}\\
qF_{\mu\nu}\!=\!-(\nabla_{\mu}P_{\nu}\!-\!\nabla_{\nu}P_{\mu})\label{Far}
\end{gather}
showing that tensorial connection and momentum are the covariant potentials of curvature and strength.

For the N-JL model, the Dirac equation (\ref{D:N-JL}), written in polar variables, becomes
\begin{gather}
\nabla_{\mu}\beta\!+\!B_{\mu}\!+\!2P^{\iota}s_{\iota}u_{\mu}
\!-\!(2P^{\iota}u_{\iota}\!-\!2m\cos{\beta}\!+\!\phi^{2})s_{\mu}\!=\!0\label{dep1:N-JL}\\
\nabla_{\mu}\ln{\phi^{2}}\!+\!R_{\mu}\!-\!2P^{\rho}u^{\nu}s^{\alpha}\varepsilon_{\mu\rho\nu\alpha}
\!+\!2ms_{\mu}\sin{\beta}\!=\!0\label{dep2:N-JL};
\end{gather}
for the S model, the Dirac equation (\ref{D:S}), in polar variables, is
\begin{gather}
\nabla_{\mu}\beta\!+\!B_{\mu}\!+\!2P^{\iota}s_{\iota}u_{\mu}\!-\!2P^{\iota}u_{\iota}s_{\mu}
\!+\!(2m\!-\!\phi^{2}\cos{\beta})s_{\mu}\cos{\beta}\!=\!0\label{dep1:S}\\
\nabla_{\mu}\ln{\phi^{2}}\!+\!R_{\mu}\!-\!2P^{\rho}u^{\nu}s^{\alpha}\varepsilon_{\mu\rho\nu\alpha}
\!+\!(2m\!-\!\phi^{2}\cos{\beta})s_{\mu}\sin{\beta}\!=\!0\label{dep2:S}:
\end{gather}
as previously announced, these are just the $p\!=\!1$ and $p\!=\!0$ cases of
\begin{gather}
\nabla_{\mu}\beta\!+\!B_{\mu}\!+\!2P^{\iota}s_{\iota}u_{\mu}\!-\!2P^{\iota}u_{\iota}s_{\mu}
\!+\!2ms_{\mu}\cos{\beta}\!-\!\phi^{2}s_{\mu}[(\cos{\beta})^{2}\!+\!p(\sin{\beta})^{2}]\!=\!0
\label{dep1:merge}\\
\nabla_{\mu}\ln{\phi^{2}}\!+\!R_{\mu}\!-\!2P^{\rho}u^{\nu}s^{\alpha}\varepsilon_{\mu\rho\nu\alpha}
\!+\!2ms_{\mu}\sin{\beta}\!+\!(p\!-\!1)\phi^{2}s_{\mu}\sin{\beta}\cos{\beta}\!=\!0
\label{dep2:merge}
\end{gather}
which is the merged N-JL/S model, that is the Dirac equation (\ref{D:merge}), in polar form. In it the auxiliary vector potentials $\varepsilon_{\mu\alpha\nu\iota}R^{\alpha\nu\iota}/2\!=\!B_{\mu}$ and $R_{\mu\nu}^{\phantom{\mu\nu}\nu}\!=\!R_{\mu}$ were introduced. By following the methods presented in \cite{Fabbri:2024avj} it is easy to check that these equations imply back the Dirac equation, and therefore they are equivalent.

The presented formulation has a number of advantages: it converts the spinor, an intrinsically complex object, into its bilinears, which are all real; it dispenses with the use of the Clifford matrices and their representation; it dispenses with the choice of tetrads. But most importantly, it converts relativistic quantum mechanics into a type of relativistic hydrodynamics with spin, where all fields are classical in essence. The treatment of the field equations, especially for the search of analytic exact solutions \cite{Fabbri:2021nkn}, becomes much easier, as we shall see in the next sections.

The last step to find explicit solutions is to introduce a parametrization for the velocity and spin vectors (\ref{SU}).
\section{A special parametrization}\label{parameter}
We will work in spherical coordinates, where the metric is given by
\begin{gather}
g_{tt}\!=\!1\ \ \ \ \ \ \ \ g_{rr}\!=\!-1\ \ \ \ \ \ \ \ g_{\theta\theta}\!=\!-r^{2}
\ \ \ \ \ \ \ \ g_{\varphi\varphi}\!=\!-r^{2}|\!\sin{\theta}|^{2}
\end{gather}
generating the connection
\begin{gather}
\Lambda^{\theta}_{\theta r}\!=\!1/r\ \ \ \ \ \ \ \
\Lambda^{r}_{\theta\theta}\!=\!-r\ \ \ \ \ \ \ \
\Lambda^{\varphi}_{\varphi r}\!=\!1/r\ \ \ \ \ \ \ \
\Lambda^{r}_{\varphi\varphi}\!=\!-r|\!\sin{\theta}|^{2}\ \ \ \ \ \ \ \
\Lambda^{\varphi}_{\varphi\theta}\!=\!\cot{\theta}\ \ \ \ \ \ \ \
\Lambda^{\theta}_{\varphi\varphi}\!=\!-\cos{\theta}\sin{\theta}\label{conn}
\end{gather}
and which has zero curvature.

On this background, we choose the specific parametrization for the velocity given by
\begin{gather}
u_{t}\!=\!\cosh{\alpha}\ \ \ \ \ \ \ \ \ \ \ \ \ \ \ \
u_{\varphi}\!=\!r\sin{\theta}\sinh{\alpha}\label{u}
\end{gather}
with $\alpha\!=\!\alpha(r,\theta)$ a generic function. The spin can then be chosen as
\begin{gather}
s_{r}\!=\!\cos{\gamma}\ \ \ \ \ \ \ \ \ \ \ \ \ \ \ \
s_{\theta}\!=\!r\sin{\gamma}\label{s}
\end{gather}
with $\gamma\!=\!\gamma(r,\theta)$ another generic function. Finally, the tensorial connection
\begin{gather}
R_{\theta\varphi\varphi}\!=\!-r^{2}\cos{\theta}\sin{\theta}\ \ \ \ \ \ \ \ \ \ \ \ \ \ \ \
R_{r\varphi\varphi}\!=\!-r|\!\sin{\theta}|^{2}\label{Rgeom}\\
R_{r\theta\theta}\!=\!-r(1\!+\!\partial_{\theta}\gamma)\ \ \ \ \ \ \ \
R_{\theta rr}\!=\!r\partial_{r}\gamma\ \ \ \ \ \ \ \
R_{t\varphi\theta}\!=\!r\sin{\theta}\partial_{\theta}\alpha\ \ \ \ \ \ \ \
R_{t\varphi r}\!=\!r\sin{\theta}\partial_{r}\alpha\label{Rgeomphys}
\end{gather}
is compatible with the parametrization (\ref{u}-\ref{s}) as clear from (\ref{ds-du}). Also, (\ref{Rie}) confirms zero curvature.

We also pick the momentum to be that of a particle of mass $m$ and spin $1/2$ as
\begin{gather}
P_{t}\!=\!m\ \ \ \ \ \ \ \ \ \ \ \ \ \ \ \ P_{\varphi}\!=\!1/2
\end{gather}
with (\ref{Far}) confirming the vanishing of the strength.

With these elements, the merged N-JL/S equations (\ref{dep1:merge}-\ref{dep2:merge}) are straightforwardly computed to be
\begin{gather}
r\partial_{r}\beta\!+\!\partial_{\theta}\alpha
\!+\!\left(-2mr\cosh{\alpha}\!+\!\frac{\sinh{\alpha}}{\sin{\theta}}
\!-\!r\phi^{2}[(\cos{\beta})^{2}\!+\!p(\sin{\beta})^{2}]\!+\!2mr\cos{\beta}\right)\cos{\gamma}
\!=\!0\label{a:merge}\\
\partial_{\theta}\beta\!-\!r\partial_{r}\alpha
\!+\!\left(-2mr\cosh{\alpha}\!+\!\frac{\sinh{\alpha}}{\sin{\theta}}
\!-\!r\phi^{2}[(\cos{\beta})^{2}\!+\!p(\sin{\beta})^{2}]\!+\!2mr\cos{\beta}\right)\sin{\gamma}
\!=\!0\label{b:merge}\\
r\partial_{r}\ln{\phi^{2}}\!+\!2
\!+\!2mr\cos{\gamma}\sin{\beta}\!+\!(p\!-\!1)r\phi^{2}\cos{\gamma}\sin{\beta}\cos{\beta}
\!+\!\partial_{\theta}\gamma
\!-\!\left(2mr\sinh{\alpha}\!-\!\frac{\cosh{\alpha}}{\sin{\theta}}\right)\sin{\gamma}\!=\!0\label{c:merge}\\
\partial_{\theta}\ln{\phi^{2}}\!+\!\cot{\theta}
\!+\!2mr\sin{\gamma}\sin{\beta}\!+\!(p\!-\!1)\phi^{2}r\sin{\gamma}\sin{\beta}\cos{\beta}
\!-\!r\partial_{r}\gamma
\!+\!\left(2mr\sinh{\alpha}\!-\!\frac{\cosh{\alpha}}{\sin{\theta}}\right)\cos{\gamma}\!=\!0\label{d:merge}
\end{gather}
which are the equations in the form that is best-suited for us in order to look for some explicit solutions.
\section{Separation of variables}\label{separated}
In the above equations, all fields depend on the latitude and radial coordinate only. It may be therefore compelling to find a situation in which these two coordinates are separated. To do this, we assume the following
\begin{gather}
\sinh{\alpha}\!=\!\frac{\sin{\theta}}{\sqrt{Z^{2}\!+\!(\cos{\theta})^{2}}}
\ \ \ \ \cosh{\alpha}\!=\!\frac{\sqrt{Z^{2}\!+\!1}}{\sqrt{Z^{2}\!+\!(\cos{\theta})^{2}}}
\ \ \ \ \ \ \ \ \sin{\gamma}\!=\!\frac{Z\sin{\theta}}{\sqrt{Z^{2}\!+\!(\cos{\theta})^{2}}}
\ \ \ \ \cos{\gamma}\!=\!\frac{\sqrt{Z^{2}\!+\!1}\cos{\theta}}{\sqrt{Z^{2}\!+\!(\cos{\theta})^{2}}}
\label{p}
\end{gather}
for the rapidity and angle, and
\begin{gather}
\sin{\beta}\!=\!-\frac{\cos{\theta}}{\sqrt{Z^{2}\!+\!(\cos{\theta})^{2}}}
\ \ \ \ \ \ \ \ \cos{\beta}\!=\!\frac{Z}{\sqrt{Z^{2}\!+\!(\cos{\theta})^{2}}}\label{ca}
\end{gather}
for the chiral angle, and where the field $Z$ is assumed to have radial dependence only. Then
\begin{gather}
\partial_{\theta}\gamma\!=\!\frac{Z\sqrt{Z^{2}\!+\!1}}{Z^{2}\!+\!(\cos{\theta})^{2}}\ \ \ \ \ \ \ \
\ \ \ \ \ \ \ \ r\partial_{r}\gamma\!=\!\frac{\cos{\theta}\sin{\theta}}{Z^{2}\!+\!(\cos{\theta})^{2}}\frac{rZ'}{\sqrt{Z^{2}\!+\!1}}\\
\partial_{\theta}\alpha\!=\!\frac{\sqrt{Z^{2}\!+\!1}\cos{\theta}}{Z^{2}\!+\!(\cos{\theta})^{2}}\ \
\ \ \ \ \ \ \ \ \ \ \ \ \ \ r\partial_{r}\alpha\!=\!-\frac{Z\sin{\theta}}{Z^{2}\!+\!(\cos{\theta})^{2}}\frac{rZ'}{\sqrt{Z^{2}\!+\!1}}
\end{gather}
as well as
\begin{gather}
\partial_{\theta}\beta\!=\!\frac{Z\sin{\theta}}{Z^{2}\!+\!(\cos{\theta})^{2}}\ \ \ \ \ \ \ \
\ \ \ \ \ \ \ \ r\partial_{r}\beta\!=\!\frac{rZ'\cos{\theta}}{Z^{2}\!+\!(\cos{\theta})^{2}}
\end{gather}
after a straightforward computation.

After re-naming $Z\!=\!\sinh{\zeta}$ and having all the above plugged into (\ref{a:merge}-\ref{d:merge}), these become
\begin{gather}
r\partial_{r}\zeta\!=\!\frac{[p(\cos{\theta})^{2}\!+\!(\sinh{\zeta})^{2}]r\phi^{2}}{\sqrt{(\sinh{\zeta})^{2}\!+\!(\cos{\theta})^{2}}}\!+\!2mr\cosh{\zeta}\!-\!2mr\sinh{\zeta}\!-\!2
\end{gather}
for the $\zeta$ function and
\begin{gather}
r\partial_{r}\ln{\phi^{2}}\!=\!\frac{(p\!-\!1)r\phi^{2}\sinh{\zeta}\cosh{\zeta}
(\cos{\theta})^{2}}{[(\sinh{\zeta})^{2}\!+\!(\cos{\theta})^{2}]^{3/2}}\!-\!2
\!+\!\frac{2mr\cosh{\zeta}(\cos{\theta})^{2}
\!+\!2mr(\sin{\theta})^{2}\sinh{\zeta}
\!-\!2\sinh{\zeta}\cosh{\zeta}}{(\sinh{\zeta})^{2}\!+\!(\cos{\theta})^{2}}
\end{gather}
\begin{gather}
\partial_{\theta}\ln{\phi^{2}}\!=\!\frac{(p\!-\!1)r\phi^{2}\sin{\theta}\cos{\theta}(\sinh{\zeta})^{2}}{[(\sinh{\zeta})^{2}\!+\!(\cos{\theta})^{2}]^{3/2}}
\!+\!\frac{(r\partial_{r}\zeta\!-\!2mr\cosh{\zeta}\!+\!2mr\sinh{\zeta}\!+\!1)\sin{\theta}\cos{\theta}}{(\sinh{\zeta})^{2}\!+\!(\cos{\theta})^{2}}
\end{gather}
for the module. Their solution is
\begin{gather}
\zeta\!=\!\ln{(2mr)}
\end{gather}
with
\begin{gather}
\phi^{2}\!=\!\frac{2\sqrt{(\sinh{\zeta})^{2}\!+\!(\cos{\theta})^{2}}}{r[(\sinh{\zeta})^{2}
\!+\!p(\cos{\theta})^{2}]}
\end{gather}
as a direct substitution would show.

The function $Z$ is given by
\begin{eqnarray}
&Z\!=\!\frac{1}{2}\left(2mr\!-\!\frac{1}{2mr}\right)\label{solution}
\end{eqnarray}
in terms of which now one could write all (\ref{p}-\ref{ca}) explicitly.

The module is also explicitly given by
\begin{gather}
\phi^{2}\!=\!\frac{8m\sqrt{\left(4m^{2}r^{2}\!-\!1\right)^{2}\!+\!(4mr\cos{\theta})^{2}}}{[\left(4m^{2}r^{2}\!-\!1\right)^{2}\!+\!p(4mr\cos{\theta})^{2}]}\label{mod:merge}
\end{gather}
and because the square root is always well-defined, the only singular points are those vanishing the denominator: as in our case $p$ is positive, this can only happen when $2mr\!=\!1$ and $p\cos{\theta}\!=\!0$ hold. This means that the singular regions are either a ring of radius $R$ such that $2mR\!=\!1$ and lying on the equatorial plane (for generic $p$) or a sphere of radius $R$ (for $p\!=\!0$). This is consistent with the fact that the Soler model only has scalar, isotropic contributions.

While writing the solution in a single form had considerable advantages, from now on we will keep the two solutions separated. So, the explicit modules for the Nambu--Jona-Lasinio ($p\!=\!1$) and Soler ($p\!=\!0$) models are
\begin{gather}
\phi^{2}\!=\!\frac{8m}{\sqrt{16m^{4}r^{4}\!+\!8m^{2}r^{2}\cos{(2\theta)}\!+\!1}}
\label{M1}
\end{gather}
and
\begin{gather}
\phi^{2}\!=\!\frac{8m\sqrt{16m^{4}r^{4}\!+\!8m^{2}r^{2}\cos{(2\theta)}\!+\!1}}{(4m^{2}r^{2}\!-\!1)^{2}}
\label{M2}
\end{gather}
respectively. The cylindrical and spherical singularities are clear.

Notice also that the distribution drops far from the origin as $1/r^{2}$ as expected for spherical waves. This means that the solution also fails to be square-integrable at infinity. We will be back to this point in the following section.
\section{Linearized equations}\label{linearized}
In the section above we found the two solutions for the nonlinear Dirac equations: the material distribution is (\ref{M1}) for the N-JL model and (\ref{M2}) for the S model. The N-JL model has ring-like and the S model has ball-like singularities.

Both distributions, for large $r$, tend to the $1/r^{2}$ spherical wave drop.

In this limit, the rest of the solution is seen to behave as
\begin{gather}
\alpha\!=\!0\ \ \ \ \ \ \ \ \gamma\!=\!\theta\label{ac}\\
\beta\!=\!0\label{b}
\end{gather}
the last of which telling us that as we move towards infinity the N-JL model would tend to isotropize into the S model.

Then, as the nonlinear terms are multiplied by an extra $\phi^{2}$ factor, they tend to disappear in the same limit.

The limit $r\!\rightarrow\!\infty$ is therefore the one for which both models linearize. It is then interesting to see what happens for the linearized case when (\ref{ac}-\ref{b}) are enforced from the start.

When equations (\ref{a:merge}-\ref{d:merge}) are linearized they become
\begin{gather}
r\partial_{r}\ln{\phi^{2}}\!+\!4\!=\!0\\
\partial_{\theta}\ln{\phi^{2}}\!=\!0
\end{gather}
in which (\ref{ac}-\ref{b}) were substituted. They can be integrated to give
\begin{gather}
\phi^{2}\!=\!\frac{K^{2}}{r^{4}}
\label{Mlin}
\end{gather}
with $K$ integration constant. This solution is square-integrable at infinity.

The fact that the solutions of the limiting linearized equation are more than the limiting solutions of the nonlinear equations is well known in the theory of differential equations and should not be surprising.
\section{Full torsion}\label{fulltorsion}
Let us consider now the field equations we would have had if we did not make any effective approximation.

The torsion field equations (\ref{torsionfe}) have now to be considered in full. We can assume torsion to have only
\begin{gather}
XW_{r}\!=\!B\ \ \ \ \ \ \ \ \ \ \ \ \ \ \ \ XW_{\theta}\!=\!C,
\end{gather}
where $B\!=\!B(r,\theta)$ and $C\!=\!C(r,\theta)$: then (\ref{torsionfe}) become
\begin{gather}
\partial_{r}\partial_{\theta}(\sin{\theta}C)
\!-\!\partial_{\theta}(\sin{\theta}\partial_{\theta}B)
\!+\!M^{2}Br^{2}\sin{\theta}\!=\!2X^{2}\phi^{2}r^{2}\sin{\theta}\cos{\gamma}\\
\partial_{r}\partial_{r}C\!-\!\partial_{\theta}\partial_{r}B
\!-\!M^{2}C\!=\!-2X^{2}\phi^{2}r\sin{\gamma}
\end{gather}
which should be considered together with the spinor equations.

Now, as it happens, we have that if expressions (\ref{p}-\ref{ca})-(\ref{solution}) and (\ref{M1}) are complemented with
\begin{gather}
B\!=\!\frac{1}{r}\frac{\sqrt{Z^{2}\!+\!1}\cos{\theta}}{Z^{2}\!+\!(\cos{\theta})^{2}}\ \ \ \ \ \ \ \
\ \ \ \ \ \ \ \ \ \ \ \ \ \ \ \ C\!=\!\frac{Z\sin{\theta}}{Z^{2}\!+\!(\cos{\theta})^{2}}
\end{gather}
then this set of fields is a solution of the full system given by (\ref{spinorfe}) and with (\ref{torsionfe}) considered in full.

Therefore, this solution is an analytic exact solution of the coupled system of field equations. It is to be remarked, however, that this fact is not trivial: if we were to consider (\ref{p}-\ref{ca})-(\ref{solution}) and now (\ref{M2}) there would be no function $H$ for which the set of fields would solve the full system (\ref{spinorfe'}) with (\ref{Higgsfe}) considered in full. Thus, the possibility to have the solution of the nonlinear equation (obtained in effective approximation) extended to a solution of the full system of equations (with no effective approximation) is a circumstance that occurs in the case of torsion, but not in the case of the Higgs field. As it stands, we do not have any physical interpretation for this.

It may be obvious, but still important, to point out that the singularity of the matter distribution has not been in any way removed, nor even slightly alleviated, by the presence of torsion. This disproves some of the results found in \cite{Fabbri:2017rjf} and references therein, claiming the absence of singularities for spinor fields with torsion.
\section{Conclusion}
In this work, we have considered the nonlinear Dirac equation in the two cases given when the nonlinearity is of a Nambu--Jona-Lasinio type and of a Soler type. We have found analytic exact solutions for both models.

The N-JL matter distribution (\ref{M1}) has a ring singularity while the S matter distribution (\ref{M2}) has a shell singularity: for both the size of the singular region is that of the Compton wave-length. Both solutions dropped as $1/r^{2}$ at infinity.

Interestingly, Takahashi discusses similar configurations of strings and balls, though only numerically \cite{Takahashi}.

We have discussed that in the limit $r\!\rightarrow\!\infty$ the N-JL model isotropizes to the S model and then both linearize.

In this case, we found another solution given by (\ref{Mlin}) dropping as $1/r^{4}$ and so square-integrable at infinity.

We found that (\ref{M1}) can be complemented to be solution for the full system of torsion and spinor equations (a fact that does not happen with (\ref{M2}) for the Higgs and spinor equations), but still displaying a ring singularity.

Finding a singularity-free spinor in interaction with torsion and possibly in nontrivial backgrounds is the next step.
\vspace{10pt}

\textbf{Funding}. This work is carried out in the framework of the INFN Research Project QGSKY and funded by Next Generation EU via the project ``Geometrical and Topological effects on Quantum Matter (GeTOnQuaM)''.

\

\textbf{Data availability}. The manuscript does not have associated data in any repository.

\

\textbf{Conflict of interest}. There is no conflict of interest.
\appendix
\section{Standard form of the solutions}
The results found in the paper rely heavily on the use of the polar decomposition. Nevertheless, some reader may find it helpful to see the solution in the usual standard form.

In the case of the merged N-JL/S model, the spinor is
\begin{gather}
\psi\!=\!\phi e^{-i(mt+\varphi/2)}e^{-\frac{i}{2}\beta\boldsymbol{\pi}}\left(\!\begin{tabular}{c}
$1$\\
$0$\\
$1$\\
$0$
\end{tabular}\!\right)
\end{gather}
with $\phi$ given by (\ref{mod:merge}), and $\beta$ given by (\ref{ca}). The above spinor is in rest-frame and spin-eigenstate, where tetrads are
\begin{gather}
\xi_{0}^{t}\!=\!\cosh{\alpha}\ \ \ \ \xi_{2}^{t}\!=\!-\sinh{\alpha}\\
\xi_{1}^{r}\!=\!\sin{\gamma}\ \ \ \ \xi_{3}^{r}\!=\!-\cos{\gamma}\\
\xi_{1}^{\theta}\!=\!-\frac{1}{r}\cos{\gamma}\ \ \ \
\xi_{3}^{\theta}\!=\!-\frac{1}{r}\sin{\gamma}\\
\xi_{0}^{\varphi}\!=\!-\frac{1}{r\sin{\theta}}\sinh{\alpha}\ \ \ \
\xi_{2}^{\varphi}\!=\!\frac{1}{r\sin{\theta}}\cosh{\alpha}
\end{gather}
generating the spin connection
\begin{eqnarray}
&C_{02r}\!=\!-\partial_{r}\alpha\ \ \ \ \ \ \ \
C_{02\theta}\!=\!-\partial_{\theta}\alpha\\
&C_{13r}\!=\!-\partial_{r}(\theta\!+\!\gamma)\ \ \ \ \ \ \ \
C_{13\theta}\!=\!-\partial_{\theta}(\theta\!+\!\gamma)\\
&C_{01\varphi}\!=\!-\cos{(\theta\!+\!\gamma)}\sinh{\alpha}\\
&C_{03\varphi}\!=\!-\sin{(\theta\!+\!\gamma)}\sinh{\alpha}\\
&C_{23\varphi}\!=\!\sin{(\theta\!+\!\gamma)}\cosh{\alpha}\\
&C_{12\varphi}\!=\!-\cos{(\theta\!+\!\gamma)}\cosh{\alpha}
\end{eqnarray}
in which $\alpha$ and $\gamma$ are given by (\ref{p}). With all these elements, we can compute the Dirac equation (\ref{D:merge}) explicitly. It is then a straightforward (albeit tedious) exercise to check that it is verified.

As for the linearized solution, it is explicitly given by
\begin{gather}
\psi\!=\!\frac{K}{r^{2}}e^{-i(mt+\varphi/2)}\left(\!\begin{tabular}{c}
$1$\\
$0$\\
$1$\\
$0$
\end{tabular}\!\right)
\end{gather}
with tetrads
\begin{gather}
\xi_{0}^{t}\!=\!1\\
\xi_{1}^{r}\!=\!\sin{\theta}\ \ \ \ \xi_{3}^{r}\!=\!-\cos{\theta}\\
\xi_{1}^{\theta}\!=\!-\frac{1}{r}\cos{\theta}\ \ \ \
\xi_{3}^{\theta}\!=\!-\frac{1}{r}\sin{\theta}\\
\xi_{2}^{\varphi}\!=\!\frac{1}{r\sin{\theta}}
\end{gather}
and it is a very easy procedure to check that this solves the free Dirac equation.

\end{document}